\documentclass[epjCONF, onecolumn]{svjour}
\usepackage{graphics}
\usepackage[varg]{txfonts} 
\usepackage[latin1]{inputenc}

\def\Msun{\,\textrm {M}_\odot}                                              
\def\kpc{\,{\rm kpc}}                                                           
\def\Gyr{\,{\rm Gyr}}                                                           
                                                           
\def\I{\textrm{\scriptsize{I}}}

\session-title{Assembling the puzzle of the Milky Way: Le Grand-Bornand, April 17-22, 2011}

\begin{document}

\title{Dynamics of stellar and H$\I$ streams in the Milky Way halo}
\author{Shoko Jin\inst{1}\fnmsep\thanks{\email{shoko@ari.uni-heidelberg.de}}\thanks{Alexander von Humboldt research fellow}}
\institute{Astronomisches Rechen-Institut, Zentrum f\"ur Astronomie der Universit\"at Heidelberg, M\"onchhofstrasse 12-14, D-69120 Heidelberg, Germany}

\abstract{
Stellar streams are key players in many aspects of Milky Way studies and, in particular, studying their orbital dynamics is crucial for furthering our understanding of the Milky Way's gravitational potential. Although this is not a trivial task when faced with incomplete dynamical phase-space information, transverse motions of streams can nevertheless be comprehended by harnessing the information contained within their radial velocity gradients.  Such methods are not only applicable to stellar streams, but also to H$\I$ streams residing in the Milky Way halo.  Here, I present the results of two studies that use radial velocity gradients to determine the system's orbit: for Hercules, one of the `ultra-faint' dwarf galaxies exhibiting a large ellipticity and located at a distance of 140~kpc, showing that it may in fact be a stellar stream, and for a string of high-velocity H$\I$ clouds belonging to the GCN complex, indicating its likelihood for being a gaseous stream at a distance of approximately 20~kpc.}

\maketitle

\section{Introduction}
\label{intro}

Stellar streams are ubiquitous in galaxy halos. Not only are they observed in abundance within our own Milky Way galaxy, they are also found to be teeming around the neighbouring Andromeda galaxy (PAndAS, \cite{2009Natur.461...66M}). The Tidal Stream Pilot Survey (\cite{2010AJ....140..962M}) has also provided an insight into the accretion of dwarf galaxies by other Local-Volume galaxies through deep, wide photometry, revealing the signatures left by the close flyby of satellite galaxies around their host galaxy. The mounting evidence from these and other observations points to hierarchical growth and evolution of Milky Way-type galaxies through the continued assimilation of their satellite systems, sometimes long after the larger galaxies have settled down in their present-day form.

Signs of disruption around nearby satellite galaxies and globular clusters, as well as more established stellar streams (often without an obvious progenitor, e.g. Local-Volume cases in \cite{2010AJ....140..962M} or the Orphan stream in the Milky Way, \cite{2007ApJ...658..337B}) leave us with tantalising pieces of information regarding the motions of stellar systems within large, massive dark-matter halos that are nevertheless insufficient for us to reconstruct the orbital path of the satellite and/or of the stream. In many cases, this is partly due to the two-dimensional sky projection being the only information available. In other cases, deep photometry may enable distances of different regions within the stream to be probed through horizontal-branch stars or variable stars that can act as `standard candles', thus allowing accurate (differential) distances to be measured along a stream so that a three-dimensional impression of the stream can be obtained. Even so, the lack of kinematic information is often a strong hinderance in the pursuit of reconstructing a stellar system's track. But why this quest for the route taken by a satellite system or of its stream? The orbits of satellite systems and streams provide one way for probing the gravitational potential of the host galaxy's dark matter halo in regions occupied by these systems, which tend to be far beyond the reaches of the host's main stellar components (disk and bulge). And because the outer regions are where the dynamical timescales are longest, it is where the remnants of tidal disruption can remain (relatively) the most untouched.

In recent years, a number of low-luminosity dwarf galaxies have been discovered around the Milky Way through SDSS data. Collectively referred to as `ultra-faint' dwarf galaxies, some of these systems (Hercules, \cite{2007ApJ...668L..43C}; Ursa Major I, \cite{2008A&A...487..103O}; Ursa Major II, \cite{2010AJ....140..138M}) exhibit large ellipticities. While they are not known to be embedded within stellar streams, their elongated forms beg the question of whether they might be experiencing Galactic tidal forces, or undergoing tidal disruption. 

For Hercules, a tentative radial velocity gradient has been observed by \cite{2009A&A...506.1147A}. Using these data and applying a method that allows the tangential velocity of a stream to be determined from radial velocity measurements along the system (\cite{2007MNRAS.378L..64J}), Section~\ref{sec:Hercules} explores whether Hercules could have had a close enough encounter with the Milky Way to make tidal disruption a viable scenario to explain its large ellipticity (\cite{2009MNRAS.400L..43J}), in other words asking whether the stellar system could in fact be dissolving into a stream. Section~\ref{sec:GCN} presents another usage of this method on a string of high-velocity neutral hydrogen (H$\I$) clouds, showing the GCN complex located towards the Galactic centre to be a possible gaseous stream (\cite{2010MNRAS.408L..85J}).

\section{An orbit for Hercules}
\label{sec:Hercules}

\begin{figure}
\begin{center}
\resizebox{0.75\textwidth}{!}{
\rotatebox{270}{
 \includegraphics{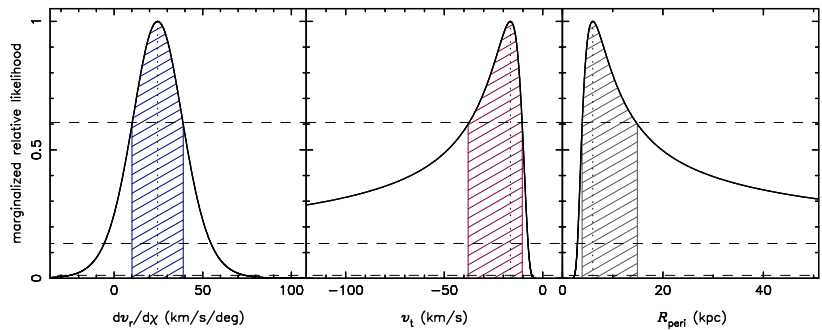}}}	
\caption{From left to right, one-dimensional likelihood distribution functions for the radial-velocity gradient, tangential velocity, and pericentric distance of the resulting orbit. The likelihood distribution function for the velocity gradient has been determined by marginalizing over the other parameters of the maximum likelihood procedure, while that of the tangential velocity has been deduced directly from applying equation~(2) of \cite{2007MNRAS.378L..64J}. The distribution function of the perigalacticon is then derived directly from orbit calculations. The dashed lines intersect the likelihood functions at the boundaries of the $1\sigma$, $2\sigma$ and $3\sigma$ confidence intervals, with the hashed regions indicating the $1\sigma$ confidence interval in each case.}	
\label{fig:MLpdfs}
\end{center}
\end{figure}	

Using the radial velocity measurements of Hercules member stars as observed by \cite{2009A&A...506.1147A}, we employ a maximum-likelihood approach in order to determine the probability distribution function of the radial velocity gradient (and from this, that of the tangential velocity) and thereby establish whether Hercules could have had a recent, close perigalactic passage. We assume a distance to Hercules of $138\pm7\kpc$, use a Galactic potential composed of a Miyamoto-Nagai disk and bulge as defined by \cite{1990ApJ...348..485P} and a Navarro-Frenk-White dark-matter halo with parameters as constrained by \cite{2008ApJ...684.1143X}, and use equation~(2) of \cite{2007MNRAS.378L..64J} to deduce the tangential (or transverse) velocity from the radial-velocity gradient. Note that using this equation gives two possible solutions, but one of these corresponds to sets of tangential velocities so large that the pericentre would be very large and the orbit of Hercules would be unbound. As this is clearly not a viable scenario for inflicting tidal disruption, this set of solutions is not considered for further analysis.

The left-most panel of Figure~\ref{fig:MLpdfs} shows the results of this analysis for the main parameter of interest, the radial-velocity gradient, from which the likelihood distribution for the tangential velocity (middle panel) can be derived. The other parameters of the maximum-likelihood procedure are the mean radial velocity of the stars and the system's internal velocity dispersion. Given the sky position, radial velocity, and the distance to and direction of motion of the assumed Hercules stream, determining the likelihood distribution of the tangential velocity is equivalent to deducing the likelihood distribution of the orbit. By integrating the orbit backward and forward in time, we obtain the likelihood distribution for the perigalactic distance (right-most panel of Figure~\ref{fig:MLpdfs}), yielding tight constraints on the small pericentric distance of $6^{+9}_{-2}\kpc$. The best model puts this flyby at $0.6\Gyr$ ago.

\section{An orbital path for the GCN stream}
\label{sec:GCN}

\begin{figure}
\begin{center}
\resizebox{0.75\textwidth}{!}{
\rotatebox{270}{
\includegraphics{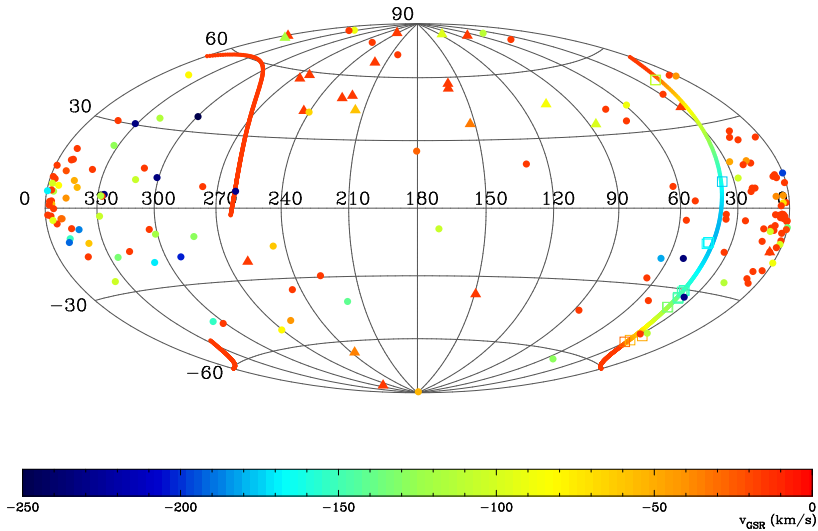}}}	
\caption{Hammer projection of the Galactic sky showing the evolution of line-of-sight velocity (with respect to the Galactic Standard of Rest) along the orbit determined for the GCN stream. Individual high-velocity clouds considered to be part of the GCN stream are indicated by open squares, whilst the solid line depicts the orbit that best matches the sky positions and velocities of the H$\I$ clouds. The positions and respective velocities of Milky Way dwarf galaxies and globular clusters are also plotted. This, along with the equivalent data for distances (see \cite{2010MNRAS.408L..85J}, Figure~4) shows that there is no obvious progenitor for the H$\I$ stream amongst the Galactic satellites.}	
\label{fig:GCN_Hammer_vel}
\end{center}
\end{figure}	

The Galactic radio sky is awash with clumps and filaments of neutral hydrogen at all perceivable velocities. Interestingly, the Milky Way also hosts many hundreds of so-called high-velocity clouds of varying sizes and morphologies with line-of-sight velocities that make them incompatible as components of the Milky Way's differentially-rotating H$\I$ disk. The most prominent high-velocity cloud structure in the Galaxy is the Magellanic stream. Gaseous structures like the Magellanic stream have well-determined kinematics, but lack the distance information more easily provided for stellar streams. Nevertheless, it is possible to analyse the dynamics of H$\I$ streams if enough radial velocity measurements sampling a sufficiently large velocity gradient are available (\cite{2007MNRAS.378L..64J,2008MNRAS.383.1686J}).

Figure~\ref{fig:GCN_Hammer_vel} shows the sky positions and line-of-sight velocities of the individual high-velocity clouds belonging to the GCN complex and that are considered to be part of a single, coherent H$\I$ stream in this analysis. By using the radial velocity gradient in this structure and following the same procedure as was described in Section~\ref{sec:Hercules}, we find an orbit (solid line in Figure~\ref{fig:GCN_Hammer_vel}) consistent with the velocities and sky positions of the H$\I$ clouds. This orbit places the stream at a heliocentric distance of approximately $10-35\kpc$ (see \cite{2010MNRAS.408L..85J}, Figure~3). A search for a possible stellar-system progenitor of the gas stream yields no obvious candidate among the currently known dwarf galaxies and globular clusters of the Milky Way, and it also does not overlap with any of the known Galactic stellar streams. The GCN stream as currently proposed therefore stands as an `orphan' H$\I$ stream of approximately $10^5\Msun$.

\section{Summary}
\label{sec:summary}

While the true nature or identity of ultra-faint dwarf galaxies remains a topic of debate, it is useful to have a means for determining whether the more distorted systems within this class could have had their stretched morphologies induced by Galactic tides. Determining the transverse velocity (i.e. proper motion) from a gradient in radial velocity offers the possibility to trace the orbital motion of a stellar system to test such a hypothesis, in particular in the current, pre-GAIA, pre-LSST era. While the application here certainly does not prove that Hercules had a close pericentric encounter with the Milky Way, it does offer the possibility of such an event in the recent past.  Likewise, the GCN stream requires further observations at lower H$\I$ column densities in between the cores of higher-density neutral hydrogen (that exhibit the smoothly varying radial velocity trend) in order for the structure to be unambiguously identified as a stream. However, the possibility of its existence as a genuine stream in the Galaxy holds exciting prospects of further discoveries of Galactic gaseous streams less massive than the Magellanic stream, especially in light of the sensitivities and resolution offered by the most recent all-sky H$\I$ surveys: GASS for the southern sky (\cite{2010A&A...521A..17K}) and EBHIS for the north (\cite{2011AN....332..637K,2011A&A...533A.105W}).

\end{document}